%Paper: hep-th/9307038
%From: WITTEN@sns.ias.edu
%Date: 06 Jul 1993 11:36:39 -0400 (EDT)
%Date (revised): 09 Jul 1993 16:10:07 -0400 (EDT)

\input harvmac.tex
%\singlespace
\def\ov{\overline}

\def\IR{{\hbox{{\rm I}\kern-.2em\hbox{\rm R}}}}
\def\IB{{\hbox{{\rm I}\kern-.2em\hbox{\rm B}}}}
\def\IN{{\hbox{{\rm I}\kern-.2em\hbox{\rm N}}}}
\def\IC{{\ \hbox{{\rm I}\kern-.6em\hbox{\bf C}}}}
\def\IP{{\hbox{{\rm I}\kern-.2em\hbox{\rm P}}}}

\def\IZ{{\hbox{{\rm Z}\kern-.4em\hbox{\rm Z}}}}
\def\QP{\IP_{4}(5)}

%comment out \def\blackb#1{{\fam\black\relax#1}}

%Then UNcomment the definition: \def\blackb{\bf}

\def\bq{\bar Q_{+,L}}

% this is the definition for an overleftrightarrow
\def\overlrarrow#1{\vbox{\ialign{##\crcr
$\leftrightarrow$\crcr\noalign{\kern-1pt\nointerlineskip}
$\hfil\displaystyle{#1}\hfil$\crcr}}}
%Use it in an equation as shown below
%$$
%\overlrarrow{\partial}
%$$

%%%%%%%%%%%%%%%%%%%%%%%%%%%%%%%%%%%%%%%%%%%%%%%%%%%%%%%%%%%%%%%%%%%%%%%%%%%
%Blackboard letters
%  The prehistoric version of this font is known as "msym". Many
%  still have this old (and UGLY) ancestor of "msbm".

\font\blackboard=msbm10 \font\blackboards=msbm7
\font\blackboardss=msbm5
\newfam\black
\textfont\black=\blackboard
\scriptfont\black=\blackboards
\scriptscriptfont\black=\blackboardss
\def\ov{\overline}
\def\bar{\overline}
%   Those who have neither "msbm not "msym" fonts can
% substitute
%   the definition

%
%
%

\lref\GSO{F. Gliozzi, J. Scherk, and D. Olive, ``Supersymmetry,
Supergravity Theories, and the Dual Spinor Model,'' {\it Nucl. Phys.}~
{\bf B122} (1977) 253.}

\lref\Fuchs{J. Fuchs, A. Klemm and M. Schmidt, ``Orbifolds by Cyclic
Permutations in Gepner Type Superstrings and in the Corresponding
Calabi-Yau Manifolds,'' {\it Ann. Phys.}~{\bf 214} (1992) 221.}
\lref\Rohm{R. Rohm and E. Witten, ``The Antisymmetric Tensor Field in
Superstring Theory,''~{\it Ann. Phys.}~{\bf 170} (1986) 454.}
\lref\Charge{X.G. Wen and E. Witten, ``Electric and Magnetic Charges in
Superstring Models,'' {\it Nucl. Phys.}~{\bf B261} (1985) 651.}
\lref\Cecotti{S. Cecotti, ``N=2 Landau-Ginzburg v.s. Calabi-Yau $\sigma$
Models: Non-Perturbative Aspects,'' {\it Int. J.  Mod.
Phys.} {\bf A6} (1991) 1749.}
\lref\Witten{E. Witten, ``Phases of N=2 Theories in Two Dimensions,''
Institute for Advanced
Study Preprint IASSNS-HEP-93/3, to appear in {\it Nucl. Phys.} {\bf B}.}
\lref\Lance{L. Dixon, ``Some World-Sheet Properties of Superstring
Compactifications on Orbifolds and Otherwise,'' in {\it Proceedings of
the 1987 ICTP Summer Workshop in High Energy Physics and Cosmology },
e.d. G. Furlan et. al.}
\lref\Martinec{E. Martinec, ``Algebraic Geometry and Effective
Lagrangians,'' {\it
Phys. Lett.}~{\bf 217B} (1989) 431, ``Criticality, Catastrophes, and
Compactifications,'' in {\it Physics and Mathematics of Strings}, ed. L.
Brink, D. Friedan, and A. Polyakov (World Scientific, 1990).}
\lref\OneMin{D. Kastor, E. Martinec, and S. Shenker, ``RG Flow in N=1
Discrete Series,'' {\it Nucl. Phys.}~{\bf B316} (1989) 590.}
\lref\VafWar{C. Vafa and N. Warner, ``Catastrophes and the Classification
of Conformal Theories,'' {\it Phys. Lett.}~{\bf 218B} (1989) 51.}
\lref\GVW{B. Greene, C. Vafa and N. Warner, ``Calabi-Yau Manifolds and
Renormalization Group Flows," {\it Nucl. Phys.}~{\bf B324} (1989)
371.}
\lref\Zam{A.B. Zamolodchikov, ``Conformal Symmetry and Multicritical
Points in Two Dimensional Quantum Field Theory,'' {\it Sov. J. Nucl.
Phys.}~{\bf 44} (1986) 529.}
\lref\Gepner{D. Gepner, ``Exactly Solvable String Compactification on Manifolds
 of
SU(N) Holonomy,'' {\it Phys. Lett.}~{\bf 199B} (1987) 380.}
\lref\Wen {M. Dine, N. Seiberg, X.G. Wen and E. Witten,
``Non-Perturbative Effects on the String World Sheet I,'' {\it Nucl.
Phys.}~{\bf B278} (1986) 769, ``Non-Perturbative Effects on the String
World Sheet II,'' {\it Nucl. Phys.}~{\bf B289} (1987) 319. }
\lref\Distler{J. Distler, ``Resurrecting (2,0) Compactifications,'' {\it
Phys. Lett.}~{\bf 188B} (1987) 431.}
\lref\VafaQ{C. Vafa, ``Quantum Symmetries of String Vacua,''
{\it Mod. Phys. Lett.}
{}~{\bf A4} (1989) 1615. }
\lref\Wit{E. Witten, ``New Issues in Manifolds of SU(3) Holonomy,'' {\it
Nucl. Phys.}~{\bf B268} (1986) 79.}
\lref\GD{J. Distler and B. Greene, ``Aspects of (0,2) String
Compactifications,'' ~{\it Nucl. Phys.}~{\bf B304} (1988) 1.}
\lref\Candelas{P. Candelas, G. Horowitz, A. Strominger and E. Witten,
``Vacuum Configurations for Superstrings," {\it Nuclear Physics}~{\bf
B258} (1985) 46.}
\lref\Schimm{R. Schimmrigk, ``A New Construction of a Three Generation
Calabi-Yau Manifold,'' {\it Phys. Lett.}~{\bf 193B} (1987)
175.}
\lref\Gep{D. Gepner, ``String Theory on Calabi-Yau Manifolds: The Three
Generation Case,'' Princeton Preprint, 1988.}
\lref\Hubsch{M. Eastwood and T. Hubsch, ``Endormorphism Valued
Cohomology and Gauge Neutral Matter,'' {\it Comm. Math. Phys.}~{\bf 132}
(1990)~383.}
\lref\Berg{P. Berglund, T. Hubsch and L. Parkes,
``Gauge Neutral Matter in Three Generation Superstring
Compactifications,''~{\it Mod. Phys. Lett.}~{\bf A5} (1990) 1485.}
\lref\WitMin{E. Witten, ``On the Landau-Ginzburg Description of N=2
Minimal Models,'' Institute for Advanced Study Preprint
IASSNS-HEP-93/10.}
\lref\Vafalg{C. Vafa, ``String Vacua and Orbifoldized LG Models,'' {\it
Mod. Phys. Lett.}~{\bf A4} (1989) 1169.}
\lref\Kenlg{K. Intriligator and C. Vafa, ``Landau-Ginzburg Orbifolds,''
{\it Nucl. Phys.}~{\bf B339} (1990) 95.}
\lref\LVW{W. Lerche, C. Vafa and N. Warner, ``Chiral Rings in N=2
Superconformal Theories,'' {\it Nucl. Phys.}~{\bf B324} (1989) 427.}
\lref\Pasquinu{S. Cecotti, L. Girardello, and A. Pasquinucci,
``Non-perturbative Aspects and Exact Results for the N=2 Landau-Ginzburg
Models,'' {\it Nucl. Phys.}~{\bf B338} (1989) 701, ``Singularity
Theory and N=2 Supersymmetry,'' {\it Int. J. Mod. Phys.} {\bf A6} (1991)
2427.}
\lref\Miron{J. Distler, B. Greene, K. Kirklin and P. Miron,
``Calculating Endomorphism Valued Cohomology: Singlet Spectrum in
Superstring Models,'' {\it Comm. Math. Phys.}~{\bf 122} (1989) 117.}
\lref\Book{T. Hubsch, {\it Calabi-Yau Manifolds: A Bestiary For
Physicists} (World Scientific, 1992).}
\lref\Bott{R. Bott and L. Tu, {\it Differential Forms In Algebraic
Topology} (Springer-Verlag, 1982).}
\lref\fre{P. Fr\'e, F. Gliozzi, M. Monteiro, and A. Piras,
``A Moduli-Dependent Lagrangian For $(2,2)$ Theories On
Calabi-Yau $n$-Folds,'' Class. Quant. Grav. {\bf 8}
(1991) 1455; P. Fr\'e, L. Girardello, A. Lerda, and P.
Soriani, ``Topological First-Order Systems With Landau-Ginzburg
Interactions,'' Nucl. Phys. {\bf B387} (1992) 333.}

\Title{IASSNS-HEP-93/40, PUPT-1397}
{\vbox{\centerline{Computing The Complete Massless Spectrum }
	\vskip2pt\centerline{Of A Landau-Ginzburg Orbifold}}}
\centerline{Shamit Kachru\footnote{$^\dagger$}
{Research supported in part by an NSF Graduate Fellowship}}
\bigskip\centerline{Joseph Henry Laboratories}
\centerline{Jadwin Hall}
\centerline{Princeton University}\centerline{Princeton, NJ 08544 }
\bigskip
\centerline{Edward Witten\footnote{$^\star$}
{Research supported in part by NSF Grant PHY92-45317}}
\bigskip\centerline{School of Natural Sciences}
\centerline{Institute for Advanced Study}
\centerline{Olden Lane}
\centerline{Princeton, NJ 08540}

\vskip .3in

We develop techniques to compute the complete massless spectrum in
heterotic string compactification
on N=2 supersymmetric Landau-Ginzburg orbifolds.
This includes not just the
familiar charged fields, but also the gauge singlets.
The number of gauge singlets can vary in the moduli space of
a given compactification and can differ from what it would be in the
large radius limit of the corresponding Calabi-Yau.
Comparison with exactly soluble Gepner models
provides a confirmation of our results at Gepner points.  Our
methods carry over
straightforwardly to $(0,2)$ Landau-Ginzburg models.

\Date{July 1993} %replace this line by \draft  for preliminary versions
	     %or specify \draftmode at some point
%\draft

\newsec{Introduction}

Landau-Ginzburg models have long been used as mean field models
of critical phenomena.  More recently it was realized that in two
dimensions, much sharper results can be extracted from them.
For instance, minimal conformal field theories can be
described as Landau-Ginzburg models as shown for bosonic theories
by Zamolodchikov \Zam\ ; this was extended
for $N=1$ supersymmetry in \OneMin\ and for $N=2$ in
\VafWar\  \Martinec\ \Pasquinu.

The $N=2$ case has many special simplifications related in part
to the non-renormalization theorems for the superpotential.  For instance,
for $N=2$ it is possible to calculate the minimal model characters
directly from the Landau-Ginzburg model \WitMin.
Also, for $N=2$, certain orbifolds of Landau-Ginzburg models
have a beautiful and unexpected relation to Calabi-Yau sigma models
\Martinec \GVW \LVW \Cecotti \Witten.  The Landau-Ginzburg model describes
a certain ``point,'' or really a certain submanifold, in the Calabi-Yau
moduli space.

The $N=2$ models also have particularly interesting physical
applications.  $N=2$ theories with the appropriate central charge
can be used to construct compactifications of the heterotic string,
and thereby to build models of particle physics, with unbroken
space-time supersymmetry.  Landau-Ginzburg
models can in particular be used to build such compactifications
-- giving specializations of Calabi-Yau models \Vafalg \Kenlg .

These specializations are technically natural, in the usual sense
of particle physics, because of enhanced symmetries (involving
twist fields; see \Witten,
\S3.4, for an explicit explanation).  They are interesting because
of calculable stringy effects (such as the enhanced
symmetries or a deviation
of the number of massless particles from what it would be in the field
theory limit).

Also, Landau-Ginzburg models are special cases
of Calabi-Yau models in which instanton corrections are turned off
(see \Witten, \S3.4).  As the instanton corrections are the usual obstruction
to forming $(0,2)$ deformations of sigma models \Wen, it would appear
likely that $(0,2)$ Landau-Ginzburg models (which are easily constructed
\Witten, \S6) have conformally invariant infrared fixed points.  This
is then an interesting case in which conformally invariant $(0,2)$ models
should be accessible for fairly detailed study.  $(0,2)$ models are of
course of considerable interest because of their use in constructing
models of particle physics with effective four dimensional gauge groups
more realistic than $E_6$.

Except for Gepner models,
which are more or less fully constructed algebraically,
most studies
of these models have focussed on the chiral primary states.
Those states enter in many beautiful constructions and among other things
determine the spectrum of massless charged particles.
However, the massless gauge singlets are not (all) determined by the
chiral primary states, and the notion of chiral primaries does not
carry over to $(0,2)$ models.  (The two facts are related: the massless
gauge singlets that do not come from chiral primaries are represented
by vertex operators that break $N=2$ or $(2,2)$ supersymmetry down to $(0,2)$.)
Our intention in this paper is to develop methods for computing the
complete massless spectrum of Landau-Ginzburg models, both $(0,2)$
and $(2,2)$ models, and including all of the gauge singlets.

In \S2 we describe the necessary facts and methods.
In \S3 we study in detail a familiar model -- the quintic.
One virtue of this model is that (at a special point in the parameter
space) the results can be compared to known results about the
corresponding Gepner model.  It should be clear, however, that our
methods carry over without essential change
to arbitrary Landau-Ginzburg models, including
$(0,2)$ models.

For Calabi-Yau manifolds, one can identify the particles
which are massless in the field theory limit by computing suitable
cohomology groups; but difficult questions then arise, in general,
of whether instanton corrections might give non-vanishing
(but exponentially small in the field theory limit) masses to some
of these states.  For Landau-Ginzburg models, however, one can argue
-- as we will do in \S2.1 -- that our results
are actually exact.  Intuitively, this is in keeping with the fact
that the Landau-Ginzburg models have no instantons.

\newsec{Background And Methods}

We will work in $N=2$
superspace with coordinates $x^{m},~\theta^{\alpha}, ~\bar
\theta^{\dot \alpha}$ (our conventions follow those of \Witten ).
In an $N=2$ superconformal theory, there are four supersymmetry charges
$Q_{+}$,~$Q_{-}$,~$\bar Q_{+}$ and $\bar Q_{-}$, where $-$ and $+$
specify left- and right-movers on the worldsheet.\foot{We will use the
terms left-moving and right-moving somewhat loosely to
describe modes that in the conformally invariant limit are left-moving
or right-moving.}
The right moving supersymmetries satisfy
\eqn\susygen{Q_{+}^{2} = \bar Q_{+}^{2} = 0, ~\{Q_{+},~\bar Q_{+}\}
= 2~ L_{0+}}
where $L_{0+}$ is the coefficient of the zero mode in the Laurent
expansion of the right moving stress-energy tensor $T_{++}$.

The worldsheet ``matter'' that we are interested in will be
chiral superfields $\Phi$.  Such fields satisfy
\eqn\chisf {[\bar D_{+},~\Phi]~=~[\bar D_{-},~\Phi]~=~0}
where $D$ and $\bar D$ are known as superspace covariant
derivatives; the complex conjugates of the $\Phi$'s are
anti-chiral fields $\bar \Phi$ that satisfy
equation \chisf ~with $\bar D \rightarrow D$.
The chiral superfields have an expansion in terms of component fields
\eqn\compex{\Phi (x,\theta) = \phi (y) + {\sqrt 2}
\theta^{\alpha}\psi_{\alpha}(y) +
\theta^{\alpha}\theta_{\alpha}F(y).}

Recall that the most general renormalizable
Lagrangian for an $N=2$ supersymmetric theory
with chiral superfields $\Phi_{i}$ and their anti-chiral conjugates $\bar
\Phi_{i}$ has the form
\eqn\nurg{L_1 = \int ~d^{2}x~d^{4}\theta ~K(\Phi,\bar \Phi)
- \int ~d\theta^{+}~d\theta^{-}~W(\Phi) -
\int ~d\bar \theta^{+}~d\bar \theta^{-
 }
{}~\bar W(\bar \Phi)}
where $K$ is called the Kahler potential (its derivatives determine the
metric on target space; the target spaces of $N=2$ models constructed
from chiral superfields are always
Kahler manifolds) and $W$ is a holomorphic function of the fields, called
the superpotential; we will choose $K$ to have the form
$K = {\bar \Phi} \Phi $ corresponding to a flat metric.
After performing the $\theta$ integrals and integrating out the auxiliary
fields, the Lagrangian becomes
\eqn\jcnon{\eqalign{L_1 = \int \sum_{i} &
{}~d^2 x~ \bigl( -\partial_\alpha\bar\phi_i\partial^\alpha
\phi_i+i\bar\psi_{-,i}(\partial_0+\partial_1)\psi_{-,i} +
i\bar\psi_+(\partial_0-\partial_1)\psi_{+,i}\bigr.\cr
&\bigl.-\sum_i\left|{\partial W\over\partial\phi_i}\right|^2-{\partial^2
W\over\partial\phi_i\partial\phi_j}\psi_{-,i}\psi_{+,j}
-{\partial^2\bar W\over\partial\bar\phi_{i}\partial\bar\phi_{j}}
\bar\psi_{+,j}\bar \psi_{-,i }.\bigr)\cr }}

The superpotential $W(\Phi_{i})$ is said to be quasi-homogeneous if
for some integers $n_{i}$ and $d$ one has
$W(\lambda^{n_{i}}\Phi_{i}) = \lambda^{d}W(\Phi_{i})$.
Such quasi-homogeneity ensures the existence
of left- and right-moving $R$-symmetries that play an important role.
The models that are believed to be related to Calabi-Yau models
are actually not Landau-Ginzburg models as introduced above but
orbifolds in which one projects onto states with integral $R$ charges.
For future use, it is convenient to set
\eqn\mildew{\alpha_i={n_i\over d}.}
The theory described by \nurg\ is believed to flow
in the infrared to a conformal field theory with central charge
\eqn\ildew{\widehat c = \sum_i(1-2\alpha_i).}

In applications in string theory, it is necessary to consider
the model formulated in four sectors -- (R,R), (NS,R), (R,NS), and
(NS,NS), where R and NS refer to Ramond and Neveu-Schwarz boundary
conditions; the two entries give the boundary conditions for left-movers
and for right-movers.  In applications to Type II superstrings,
one would have (in models of this particular type) space-time supersymmetries
coming from both left- and right-movers.  These supersymmetries
determine the spectrum in all four sectors in terms of the spectrum
in, say, the (R,R) sector.  In practice, this means that to identify
massless particles in space-time, it suffices to find the (R,R) ground
states.  These have very special properties which have been much
exploited in the literature on Landau-Ginzburg models and their
applications.   Their (NS,NS) cousins are represented by vertex
operators that preserve (2,2) world-sheet supersymmetry.

We are actually interested in using the same models to describe
compactifications of the heterotic string.  In this case, we supplement
\nurg ~by ten left-moving free fermions
\eqn\jdnd{L_2=\int d^2x\,\,\,\sum_{I=1}^{10}\lambda_I i(\partial_0+\partial_1)
\lambda_I}
and extra degrees of freedom representing an additional $E_8$ current
algebra.  The $\lambda_I$ are given the same NS or R boundary conditions
as the left-moving part of \nurg.  The combined Lagrangian $L_1+L_2$
is expected (as in Calabi-Yau compactification) to give an unbroken
$E_6$ gauge group in space-time.

Space-time supersymmetries are now derived from right-movers only.
Therefore, there are two sectors that must be studied -- (R,R) and
(NS,R).  The study of the (NS,R) model is one of the main novelties
in this paper.  We are no longer interested only in states with a simple
relation to (R,R) ground states, so new methods must be developed.

In fact, in the (NS,R) sector, there are massless gauge singlet
states that are represented by vertex operators that (even if one
suppresses the $\lambda$'s) break $(2,2)$ world-sheet supersymmetry down to
$(0,2)$ supersymmetry.  These are the modes that, in compactification
on a Calabi-Yau
manifold $X$, arise from $H^1(X,{\rm End}(T))$. \foot{
For some computations of this cohomology group in Calabi-Yau
models see \Miron \Hubsch \Berg \Book .}
Understanding these modes in the context of Landau-Ginzburg models
is one of our main goals in this paper.  In the process of doing
this, we will automatically develop the techniques needed to
compute the complete massless spectrum in more general (0,2) Landau-Ginzburg
models.

An $SO(10)$ symmetry acting on the ten $\lambda$'s is manifest in
the above Lagrangian.  $SO(10)$ is not a maximal subgroup of $E_6$,
which instead contains an $SO(10)\times U(1)$ factor.  The $U(1)$ generator
is simply the left-moving $R$-current -- call it $J_L$ -- of the
Landau-Ginzburg theory with Lagrangian $L_1$.  The rest of $E_6$ is
harder to see explicitly; the additional currents are twist fields
coming from states in the left-moving Ramond sector.

\subsec{The Born-Oppenheimer Approximation}

Because we are looking for massless states in space-time, we can set the
space-time momentum to zero and look for worldsheet wavefunctions which
have only polynomial
dependence on the lowest oscillator modes.  In sectors with negative
vacuum energy, we have to keep the lowest excited modes of the various
fields.  This truncation of the theory to a small finite number of
modes, a worldsheet ``Born-Oppenheimer'' approximation, has been applied
before in
a string theory context in \GD\ and \Rohm.  However, the focus there
was on sigma models.  In the Landau-Ginzburg context, it is easy
to be more explicit.

What is the degree of validity of the Born-Oppenheimer approximation?
We will argue that for identifying the massless modes it is exact.

We will denote the right- and left-moving world-sheet Hamiltonians
as $L_{0+}$ and $L_{0-}$.
In the (R,R) and (NS,R) sectors that we will study, physical states
have $L_{0+}=0$; for massless particles on-shell,
the ``space-time'' part of the string does not contribute to $L_{0+}$,
so we can consider $L_{0+}$ to be the right-moving Hamiltonian of
the ``internal'' theory only.
In a right-moving Ramond sector, there
are two right-moving global supersymmetries, say $Q_+$ and $\ov Q_+$,
with
\eqn\imporel{\{Q_+,\ov Q_+\}=2{L_{0+}}, ~~~~~~~~Q_+{}^2=\ov Q_+{}^2=0.}
As in Hodge theory, it follows that the kernel of ${L_{0+}}$ is the same
as the cohomology of $\ov Q_+$.

This simple fact is the starting point for all our computations: we identify
the massless states with the cohomology of $\ov Q_+$ (or actually the subspace
of that cohomology consisting of states
with the correct eigenvalue of $L_{0-}$).  This is a great
advantage because -- due to the simple
properties of triangular matrices -- cohomology is usually highly
computable.

In the particular case at hand, the simplification comes mostly
because the $\bar Q_+$ cohomology is naturally invariant under
a rescaling of the superpotential by $W\to \epsilon W$.\foot{To be more
precise, under $W\to \epsilon W$, the $\bar Q_+$ cohomology group of
right-moving $U(1)$ charge $n$ is multiplied by $\epsilon^n$, because
of the scaling introduced momentarily.}
The reason for this is that, up to a rescaling of the fields
by
\eqn\cornox{\Phi_i\to \epsilon^{-\alpha_i}\Phi_i,}
$W\to\epsilon W$ is equivalent to a certain modification of the
kinetic energy.  The whole kinetic energy is of the form $\{\bar Q_+,\dots\}$
so the modification of the kinetic energy induced by the transformation
\cornox\ does not affect the $\bar Q_+$ cohomology.  This means that
in computing the $\bar Q_+$ cohomology, we can set $W$ to zero except
when it is needed to lift degeneracies that are otherwise present.
That fact is the basis for all of our calculations.

It is straightforward to write down the $\bar Q_+$ operator of the
Landau-Ginzburg model:
\eqn\jonio{\bar Q_+ = i{\sqrt 2}\int ~d x^1~\left(i\bar\psi_{+,i}
(\partial_0+\partial_1)\phi_{i}
+{{\partial W}\over{\partial \phi_{i}}}\psi_{-,i}
\right)}
An additional simplification arises (as in \WitMin)
because of the principle stated
in the last paragraph.
Taking $W\to \epsilon W$ and trying to compute
the $\bar Q_+$ cohomology perturbatively in $\epsilon$, the first
step is to compute the cohomology of the part of $\bar Q_+$ that is independent
of $W$:
\eqn\eening{\bar Q_{+,R}=i{\sqrt 2}\int ~d x^1~\left(i\bar\psi_{+,i}
(\partial_0+\partial_1)\phi_i\right)~.}
The cohomology of this operator is the subspace of the full Hilbert
space consisting of states in which the right-moving oscillators are
all in their ground states and which depend holomorphically
on the zero modes of the $\phi_i$; moreover the zero modes of $\psi_+$
and $\bar\psi_+$ can be omitted.  This leaves a smaller Hilbert space,
consisting of left-moving oscillators, zero modes of $\psi_-$ and
$\bar\psi_-$, and holomorphic functions of boson zero modes.  Let us
call this the left-moving Hilbert space ${\cal H}_L$.

The next step, analogous
to degenerate perturbation theory in quantum mechanics, is to compute
the cohomology of the ``perturbation''
\eqn\screening{\bar Q_{+,L}=i{\sqrt 2}\int ~d x^1~
{{\partial W}\over{\partial \phi_{i}}}
\psi_{-,i}
}
in ${\cal H}_L$.  In quantum mechanics this would usually be only the
beginning of a systematic expansion; but in the present situation we are
actually at this stage finished (at least to all finite orders),
because of the triangular nature
of cohomology and the simplicity of the cohomology of the $\bar Q_+$
operator. The requisite argument is a standard
``zig-zag'' argument, as in \Bott, p. 95, using the following facts.
Let $U$ be the operator that assigns the value $1$ to $\bar \psi_{+,i}$,
$-1$ to $\psi_{+,i}$, and 0 to other fields.  Then $[U,\bar Q_{+,R}]
=\bar Q_{+,R}$,
$[U,\bar Q_{+,L}]=0$, and the cohomology of $\bar Q_{+,R}$ is zero except at
one value of $U$.

Let us use these facts to prove that the $\bar Q_+$ cohomology is naturally
isomorphic to the cohomology of $\bar Q_{+,L}$ in the $\bar Q_{+,R}$
cohomology (which is isomorphic to ${\cal H}_L$).  So to begin with
we have a state $|\alpha_0\rangle$ that is annihilated by $\bar Q_{+,R}$
and annihilated by $\bar Q_{+,L}$ modulo $\bar Q_{+,R}(\dots)$.
We can assume that $|\alpha_0\rangle$ has $U=0$
since the $\bar Q_{+,R}$ cohomology
vanishes for other values of $U$.  The fact that $|\alpha_0\rangle$
is annihilated
by $\bar Q_{+,L}$ modulo $\bar Q_{+,R}(\dots)$ means that there
is some $|\alpha_{-1}\rangle$, necessarily of $U=-1$, such that
\eqn\guff{\bar Q_{+,L}
|\alpha_0\rangle=-\bar Q_{+,R}|\alpha_{-1}\rangle ~ .}
Then $\bar Q_+(|\alpha_0\rangle +|\alpha_{-1}\rangle)
=(\bar Q_{+,R}+\bar Q_{+,L})(|\alpha_0\rangle + |\alpha_{-1}\rangle)=\bar
Q_{+,L
 }
|\alpha_{-1}\rangle $.  Moreover
\eqn\another{\bar Q_{+,R}(\bar Q_{+,L}|\alpha_{-1}\rangle)=-\bar Q_{+,L}\bar
Q_{
 +,R}
|\alpha_{-1}\rangle=\bar Q_{+,L}\bar Q_{+,L}|\alpha_0\rangle = 0}
where the first step uses $\{\bar Q_{+,R},\bar Q_{+,L}\}=0$, the
second step uses \guff, and the last step uses $\bar Q_{+,L}{}^2=0$.
$\bar Q_{+,L}|\alpha_{-1}\rangle$ therefore represents a state in the
cohomology
of $\bar Q_{+,R}$ at $U=-1$; since the $\bar Q_{+,R}$ cohomology vanishes
except at $U=0$, this state is cohomologically trivial and there is
a state $|\alpha_{-2}\rangle$ of $ U=-2$
such that $\bar Q_{+,R}|\alpha_{-2}\rangle=-\bar Q_{+,L}|\alpha_{-1}\rangle$.
Continuing in this way, one inductively solves the equations
\eqn\solveq{\bar Q_{+,R}|\alpha_{-n-1}\rangle=-\bar
Q_{+,L}|\alpha_{-n}\rangle.}
The sum $|\alpha\rangle = |\alpha_0\rangle + |\alpha_{-1}\rangle +
|\alpha_{-2}\rangle + \dots$ is then
the desired state annihilated by $\bar Q_{+}=\bar Q_{+,R}+\bar Q_{+,L}$.
In defining $|\alpha\rangle$ and obeying the equations up to the first
$n$ terms  we have shown that the state which has zero energy in
the Born-Oppenheimer approximation has zero energy up to $n^{th}$ order
in perturbation theory in the superpotential $W$.

The question of whether the series converges is more subtle, but intuitively
this should follow from the super-renormalizability of the Landau-Ginzburg
model.  The state $\alpha_{-n}$ has $U=-n$, and as $U$ is carried only
by fermions, $\alpha_{-n}$ is
a state with very high energy, roughly at least the energy of
a degenerate fermi gas with fermi energy $n$.  For such
high energy states, $\bar Q_{+,R}$ dominates over $\bar Q_{+,L}$
because of being constructed from a current of higher dimension
(containing an extra derivative), and in the relation
\solveq, it should be possible to choose $\alpha_{-n-1}$
to be much smaller than $\alpha_{-n}$ in norm, ensuring convergence
of the series.  A rigorous proof of this assertion would be interesting.

The $\bar Q_+$ cohomology can be decomposed according to the action
of certain operators that commute with $\bar Q_+$ or have simple commutation
relations with it.  In fact, $\bar Q_+$
commutes
with the left-moving $U(1)$ charge but raises the right-moving $U(1)$
charge by one unit.\foot{The statement that $\bq$ ${\it raises}$ the
right $U(1)$ charge is convention-dependent. Our conventions for $U(1)$
charges are given in \S2.2~.}
$\bar Q_+$ also obviously commutes with the
$\lambda$'s, so states can be labeled by the number of $\lambda$ oscillators.

Somewhat less obviously \WitMin, in the Landau-Ginzburg theory \nurg, one
can find an $N=2$ superconformal algebra of left-moving fields
that commute with $\bar Q_+$.  In components, one has
\eqn\omico{\eqalign{
  J_{L} & = \sum_i\left((\alpha_i-1)\psi_{-,i}\bar\psi_{-,i}+i\alpha_i\phi_i
(\partial_0-\partial_1)\bar\phi_i\right) \cr
 G  & = -i\sqrt 2\sum_i\psi_{-,i}(\partial_0-\partial_1)\bar\phi_i \cr
 \bar G & = i\sqrt 2\sum_i\left((1-\alpha)(\partial_0-\partial_1)\phi_i
\cdot \bar\psi_{-,i}-\alpha_i\phi_i(\partial_0-\partial_1)
\bar\psi_{-,i}\right)\cr
  T & = \sum_i ~\left|(\partial_0 - \partial_1)\phi_i\right|^2
+{i\over 2}(\psi_{-,i}(\overlrarrow{\partial_0}-\overlrarrow{\partial_1})
\bar\psi_{-,i} )\cr
  &~~~~~+ {\alpha_i\over 2}\left(\partial_0-\partial_1)(i\psi_{-,i}
\bar\psi_{-,i}
-\phi_i(\partial_0-\partial_1)\bar\phi_i\right)\cr}}
In \omico, $\phi_i$, $\psi_{-,i}$, etc., are components
in the expansion \compex\ of the superfields $\Phi_i$.
Hopefully, these operators converge in the infrared to the
left-moving $N=2$ algebra of the expected
conformally invariant fixed point theory.  The central charge of
the $N=2$ algebra \omico ~is given by \ildew.  With a fairly
obvious change (renaming $\phi$ and $\partial\bar\phi$
as $\beta$ and $\gamma$) this realization of the $N=2$
algebra was first given in [\fre], where the $\bar Q_{+,L}$
operator also appeared, with a somewhat different
rationale.

There are several reasons that it is convenient to have these operators.
First of all, physical states, in addition to being annihilated
by $L_{0+}$, must have the appropriate eigenvalue of $L_{0-}$.
So among other things, we need to be able to compute the $L_{0-}$
quantum number of the Fock ground state in each sector of Hilbert space.

Furthermore, to know which $SO(10)$ singlet states are $E_6$ singlets, which
belong to $\bf{27}$'s of $E_6$, and which to $\ov {\bf{27}}$'s, we need to work
out the $J_{L}$ quantum numbers, so in particular we need to compute
the $J_{L}$ charge of the Fock ground state.  We will return to these
matters later.

A subtler reason for needing \omico ~is as follows.
In compactification on
a Calabi-Yau manifold $X$, massless gauge singlets
of the (NS,R) sector are of three kinds: states that come from
$H^1(X,T)$, states that come from $H^1(X, T^*)$, and states that
come from $H^1(X,{\rm End}(T))$.  We would like the analogous
decomposition in the case of Landau-Ginzburg models.  This can
be done as follows.  In the Calabi-Yau case, the three kinds of
states can be described as states that are annihilated by $G_{-1/2}$,
states that are annihilated by $\bar G_{-1/2}$, and states that
are annihilated by neither.
Since from \omico\ we can get an explicit and practical construction
of $G_{-1/2}$ and $\bar G_{-1/2}$, we can make the decomposition into
$H^1(X,T)$, $H^1(X,T^*)$, and $H^1(X,{\rm End}(T))$ also in the
Landau-Ginzburg case.

In addition to being
of intrinsic interest, this decomposition can be of practical use
in the following sense.
The singlets coming from $H^{1}(X,T)$ and
$H^{1}(X,T^{*})$ are in one to one correspondence with $\bf {10}$'s of
SO(10) which arise in the same twisted sectors.  The concrete form of
the correspondence is as follows.  Consider a singlet which is created
by a left chiral field,
so its representative $|\Psi\rangle$ in the $\bar Q_{+,L}$ cohomology
satisfies ${\bar G}_{-1/2}|\Psi\rangle = 0$.  Then the corresponding
$\bf {10}$
of $SO(10)$ is given by $\lambda^{I}_{-1/2}~G_{1/2}|\Psi\rangle$.
A similar construction applies to left anti-chiral singlets, with the role of
$G$ and $\bar G$ reversed.
We will
illustrate
this explicitly in the example of \S3.

\subsec {Symmetries And Quantum Numbers}

Consider an $N=2$ Landau-Ginzburg theory with chiral superfields
$\Phi_{i}$ and quasi-homogeneous superpotential $W$ such that
\eqn\quasihom{W(\lambda^{n_{i}}\Phi_{i}) ~=~\lambda^{d}~W(\Phi_{i})~}
and again set $\alpha_i=n_i/d$.
The superpotential $W$ will then have left- and right-moving charges
$(1,1)$ -- as befits a marginal operator -- if the superfields
$\Phi_i$ have charges $(\alpha_i,\alpha_i)$.\foot{In fact, the
${\it signs}$ of both $U(1)$ charges are mere conventions.  Flipping the
convention for one leads to an exchange of $\bf{27}$'s and
${\ov {\bf{27}}}$'s; this
simple observation motivated the discovery of mirror symmetry.}
In components the charges are therefore as in Table 1.
$$\vbox {\settabs 3 \columns
\+& ${\bf Table ~1}$&\cr
\+&&\cr
\+{$\bf Field$} &{$\bf q_{-}$}&{$\bf q_{+}$}\cr
\+~~{${\phi_{i}}$}&${\alpha_{i}}$&${\alpha_{i}}$ \cr
\+~~{$\bar \phi_{i}$}&$-{\alpha_i}$&$-{\alpha_i}$ \cr
\+~~$\psi^{i}_{-}$ &${{\alpha_i}-1}$&${\alpha_i}$ \cr
\+~~$\psi^{i}_{+}$ &$\alpha_i$&${\alpha_i}-1$ \cr
\+~~$\bar \psi^{i}_{-}$ &$1-{\alpha_i}$&$-{\alpha_i}$ \cr
\+~~$\bar \psi^{i}_{+}$ &$-{\alpha_i}$&$1-{\alpha_i}$ \cr}$$
\bigskip

At this point, the attentive reader might worry about the following
point.
The $J_L$ operator that transforms the fields according to the charges
given in the table is
\eqn\omiggo{J_L=\sum_i\int ~d x^1\left((\alpha_i - 1)\psi_{-,i}\bar\psi_{-,i}
+i\alpha_i\phi_i\overlrarrow{\partial_0}\bar\phi_i
+\alpha_i\psi_{+,i}\bar\psi_{+,i} \right).}
The density that is being integrated in \omiggo ~does not
commute with $\bar Q_+$, but the integrated expression does.
On the other hand, in equation \omico ~we have written down
a left-moving $U(1)$ charge density that does commute with $\bar Q_+$.
Using this density, we have a second candidate for the left-moving
$U(1)$ charge, namely
\eqn\jdnln{J'{}_L=\sum_i \int ~dx^1\left((\alpha_i-1)\psi_{-,i}\bar\psi_{-,i}
+i\alpha_i\phi_i(\partial_0-\partial_1)\bar\phi_i\right). }
Using the commutation relations
\eqn\comrules{\eqalign{\{\bar Q_+,\psi_{+,i}\} & =-\sqrt 2(\partial_0+
\partial_1)\phi_i\cr
[\bar Q_+,\bar\phi_i]& =i\sqrt 2\bar\psi_{+,i} \cr
\{\bar Q_+,\bar\psi_{-,i}\}& =i\sqrt 2 {\partial W\over\partial\phi^i}
                \cr}}
(with other components vanishing), one finds that
\eqn\snsnns{J_L=J'_L +\left\{\bar Q_+,{i\over \sqrt 2}
 \int ~dx^1 ~\left(\sum_i\alpha_i\bar\phi_i\psi_{+,i}
    \right)\right\}.}
This shows that as regards the action on the $\bar Q_+$ cohomology,
it does not matter whether we use $J_L$ or $J'_L$.  $J'_L$ arises
naturally in the simplest description of the $N=2$ algebra that acts
on the cohomology, while $J_L$ is distinguished
because it generates a symmetry even before taking the $\bar Q_+$
cohomology.

A similar question, which we might as well dispose of now,
arises for the left-moving energy operator $L_{0-}$.
The Landau-Ginzburg theory \nurg, even away from criticality,
has a conserved Hamiltonian $H$ and momentum $P$.  The conventional
$L_{0-}$ operator would be $L_{0-}=H-P$ or concretely
\eqn\standardlo{\eqalign{
L_{0-} = \int~dx^{1} &\left( ~|(\partial_{0}-\partial_{1})\phi_{i}|^{2}
- i \psi_{-,i}\overlrarrow{\partial_{1}}\bar\psi_{-,i}
\right.\cr
&\left.
 + {\partial^{2}W\over {\partial\phi_{i}\partial\phi_{j}}}\psi_{-,i}\psi_{+,j}
+ {\partial^{2}\bar W\over {\partial \bar \phi_{i}\partial \bar
\phi_{j}}}\bar\psi_{+,j}\bar\psi_{-,i}+ \left|{\partial W\over \partial
\phi_{i}}\right|^{2}~\right).}}
The $L_{0-}$ operator that we would form from the stress tensor
in \omico ~is instead
\eqn\snkn{\eqalign{L_{0-}' =
\int~dx^{1}~~&\left(|(\partial_{0} - \partial_{1} )\phi_{i}|^{2} + {i\over 2}
\psi_{-,i}{(\overlrarrow{\partial_{0}}-\overlrarrow{\partial_{1}})}
\bar\psi_{-,i} \right. \cr
&\left. + {\alpha_{i}\over 2}
(\partial_{0}-\partial_{1})(i\psi_{-,i}\bar\psi_{-,i} -
\phi_{i}(\partial_{0}-\partial_{1})\bar\phi_{i})~\right). \cr}}
In fact, $L_{0-}'=L_{0-}+\{\bar Q,\dots\}$, though a slightly
lengthy calculation is needed to show this.  For instance, to reduce \snkn
{}~to a more recognizable form, one first
writes $(\partial_0 - \partial_1)\psi_{-,i}=(\partial_0 + \partial_1)\psi_{-,i}
-2\partial_1\psi_{-,i}$, and then evaluates $(\partial_0 + \partial_1)
\psi_{-,i}$
via the equations of motion.  $(\partial_0 - \partial_1)\bar\psi_{-,i}$
can be treated similarly.  Discarding a total derivative, the term
$(\partial_0
 - \partial_1)\left(i\alpha_i\psi_{-,i}\bar\psi_{-,i}-\phi_i(\partial_0 -
\partial_1)
\bar\phi_i)\right)$ in \snkn\ can be replaced by
\eqn\mukk{(\partial_0
 + \partial_1)\left(i\alpha_i\psi_{-,i}\bar\psi_{-,i}-\phi_i(\partial_0
- \partial_1)
\bar\phi_i)\right).}
Using the fact that $\partial_0 + \partial_1\sim \{Q_+,\bar Q_+\}$,
it follows that if $\bar Q_+X=0$, then $(\partial_0+\partial_1)X=
\{\bar Q_+,\dots\}$.  Applying this principle with $X$ being
the current in the first line in \omico, we find that up to $\{\bar Q_+,\dots
\}$, \mukk\ can be replaced by $(\partial_0+\partial_1)(i\psi_{-,i}\bar
\psi_{-,i})$.  This in turn can be evaluated using the equations of motion.
After adding one last correction term
\eqn\jukk{\left\{\bar Q_+,{-{i\over \sqrt 2}}\int ~d x^1 {\partial
\bar W\over\partial
\bar\phi^i}\bar\psi_{-,i}\right\} }
to \snkn\ one obtains the desired results that $L_{0-}=L_{0-}'$
modulo $\{\bar Q_+,\dots\}$.
The significance of this is similar to the significance of the analogous
statement demonstrated for the currents in the last paragraph:
$L_{0-}'$ is more closely related to the $N=2$ algebra that acts
on the cohomology, but $L_{0-}$ is natural because it generates
a symmetry even before taking the cohomology.

The equivalence of the two $J_L$ operators and of the two $L_{0-}$ operators
means that the ground state quantum numbers
are independent of $W$ (which does not appear in
$J_L'$ and $L_{0-}'$) and can be computed using the standard formulas
associated with normal-ordering of $J_L$ and $L_{0-}$.

\subsec{Construction Of The Orbifold}

Calabi-Yau sigma models are related not quite to Landau-Ginzburg models
but to certain Landau-Ginzburg orbifolds.  These are orbifolds
in which one projects on integral values of $J_L$; $J_R$ then automatically
also becomes integral.
The projection is made
by dividing by the group generated by
\eqn\projector{e^{-{2\pi i \oint J_L(z)}}~=~e^{-{2\pi i J_L}}}
with a due modification which we will now explain when certain
fermion zero modes are present.

In physical applications of the Landau-Ginzburg orbifold, one wishes
to sum over left-moving Ramond and Neveu-Schwarz sectors.  (This
is the GSO-like projection that enters in constructing $E_8$ current
algebra.)  In $N=2$ models, the GSO projection \GSO ~can be interpreted
as a projection onto states for which $J_L$ is even.  We are not
quite dealing here with an $N=2$ model but with a $(0,2)$ model
containing also the left-moving free fermions $\lambda_I$.
Hence, in the left-moving NS sectors, the
GSO projection that we want is the one that projects onto states
in which $J_L$ plus the number of $\lambda_I$ excitations is even.  So
we project onto states with $g=1$ where
\eqn\cnnn{g=\exp(-i\pi J_L)\cdot (-1)^\lambda .}

The necessary statement in R sectors is more subtle because of
fermion zero modes.  Let $q_-$ and $q_+$
be the left-moving and right-moving $U(1)$ charges of the ``internal''
Landau-Ginzburg theory.
Then in left-moving Ramond sectors, the GSO projection
(on states that are in the ground state of the $SO(10)$ sector) can
be summarized by saying that the value of $q_{-}$
determines whether states transform in the $\bf{16}$ or
the $\overline{\bf{16}}$ of
SO(10).  One (standard) way to understand this in more detail
is to organize the ten $SO(10)$ fermions of \jdnd ~into five complex
fermions
\eqn\complex{\eta_{I} = {1\over \sqrt 2} ~(\lambda_{2I-1} +
i \lambda_{2I})}
where $I=1,\cdots 5$.
The complex fermi fields have zero modes $\eta_{0,I}$ and
$\eta^{*}_{0,I}$ which satisfy the
standard anti-commutation relations
\eqn\anticomm{\{\eta_{0,I},\eta_{0,J} \}=
\{\eta^{*}_{0,I},\eta^{*}_{0,J}\}
= 0,~~~\{\eta_{0,I},\eta^{*}_{0,J}\} = \delta_{IJ}~. }
Then acting on the Fock vacuum $|0\rangle$ which satisfies
$\eta_{0,I}|0\rangle = 0$, a 32 dimensional
representation of $SO(10)$ is furnished by the 32 states
\eqn\etastate{\eta^{*}_{0,j_{1}}\cdots\eta^{*}_{0,j_{k}}|0\rangle ~.}
It is well known that this is a reducible representation of $SO(10)$
which decomposes into two 16 dimensional irreducible representations,
the $\bf {16}$ and the $\bf{\bar{16}}$; the $\bf{16}$ is composed of the
states in \etastate\
with $k$ even, while the $\bf {\bar
{16}}$ is given by the states in \etastate\ with $k$ odd.
Notice from \cnnn\ that the gauge fermions should be thought of
as carrying an extra $U(1)$ charge of 1, for the purposes of the
projection onto even left-moving $U(1)$ charge.  Then the states in
\etastate\ with a given value of $k$
carry a left $U(1)$ charge of $-{5\over 2} + k$
(the $-{5\over 2}$ being the charge of the Fock vacuum
$|0\rangle$; see \S2.4).  The conclusion is that
states with $q_{-} - {5\over 2}$ $\it even$
are projected onto $\bf {16}$'s of $SO(10)$, while states with
$q_{-} - {5\over 2}$ $\it odd$ are associated with
$\bf {\bar {16}}$'s of $SO(10)$.

Physical applications also involve a right-moving GSO projection,
onto states with the appropriate mod 2 right-moving fermion number.
We will be interested in
massless states, which  are always right-moving ground
states; for such states the GSO projection in right-moving
Ramond sectors means the following.  States with
$q_{+} + {3\over 2}$ even give {\it left-handed} spin one-half
massless fermions in space-time; states
with $q_{+} + {3\over 2}$ odd give {\it right-handed} ones.  The
detailed explanation involves exactly the same sort of
reasoning that we have just carried out for left-movers.
(The description of the right-moving GSO projection in right moving NS
sectors is standard but we need not give it here as we only consider
right moving R sectors in this paper.)

Since in constructing the spectrum, we project onto states with a
particular eigenvalue of the operator $g$ of equation \cnnn, modular
invariance forces us to add twisted sectors
constructed with twists by arbitrary powers of $g$.
The operator $g$ is a version of the $(-1)^F$ operator that counts
fermions modulo two.
So, starting with the completely untwisted (R,R) sector, a twist
by an even power of $g$ makes
a left-moving Ramond sector; a twist by an odd power makes a left-moving
Neveu-Schwarz sectors.  With $d$ being the least common denominator
of the charges of the $\Phi_i$,
$g^{2d}=1$, so there are $2d$ sectors twisted by $1,g,~g^{2},\dots,~g^{2d-1}$.

\subsec{Ground State Quantum Numbers}

As is well known in analogous computations, one of the main steps
in determining the spectrum of one of these models is to determine
the quantum numbers of the ground state in each twisted sector.
To be precise, in the sector twisted by $g^k$, we wish to determine
the left- and right-moving $U(1)$ charges (i.e., $J_L$ and $J_R$ eigenvalues),
and the left-moving energy ($L_{0-}$ eigenvalue) of
the ground state.  We will always consider right-moving Ramond sectors,
so the $L_{0+}$ eigenvalue of the ground state will always be zero.

First, we determine the $U(1)$ charges.
Our viewpoint is that of \Charge : the reason the twisted sectors have
fractional $U(1)$ charges is that when the fermions satisfy twisted boundary
conditions, the vacuum has a fractional fermion number.
Formally, the charge carried by a
filled fermi sea with fermions of charge $e$ is
 \eqn\chargein{Q~=~e \int_{-\infty}^{0}~dE~\rho (E)}
where $\rho (E)$ is the density of states.  This is of course divergent,
and must be regulated.  Since we are really interested in the change in
$Q$ as a
function of the twisted boundary conditions on the fermions, we can
subtract an (infinite) constant
${e\over 2}~\int_{-\infty}^{\infty}~dE~\rho (E)$
without doing any harm; we also
introduce a convergence factor:
\eqn\charge{Q~=~ -{1\over 2}~lim_{s \rightarrow 0}~
\int_{-\infty}^{\infty}~dE~sign(E)~\rho (E)~
e^{-s|E|}~.}

For our case of interest, which is left moving fermions on a circle of
circumference $2\pi$ (and coordinate $0 \leq \sigma < 2\pi$) with
Hamiltonian $-i{\partial \over {\partial \sigma}}$, the integral in
\charge\  is easily evaluated for arbitrary choice of boundary
conditions.  In particular, for fermions with boundary conditions
 \eqn\fermibc{\psi (\sigma + 2\pi) =  e^{-i\theta}\psi (\sigma)}
with $0\leq \theta < 2\pi$, one finds
 \eqn\cferm{Q~=~{e\theta \over 2\pi}~-{e\over 2} }
(so the vacuum has a
fractional fermion number of ${{\theta - \pi } \over 2\pi}$).
The above formula is valid for $0<\theta<2\pi$.  It becomes valid
for all $\theta$ after the obvious modification to
\eqn\juggo{Q~=~e\left({\theta \over 2\pi} - \left[
{\theta \over 2\pi}\right]~-{1\over 2}\right)}
where $[x]$ denotes the greatest integer less than $x$.
There is an important subtlety here.  The expression $[\theta/2\pi]$ has
a discontinuity when $\theta$ is an integral multiple of $2\pi$.
At such values of $\theta$, both values of $Q$ should be kept.
The reason for this is that precisely when $\theta=2\pi n$, with integer $n$,
there are fermion zero modes; upon quantizing them, one finds (for a single
complex fermion) a pair of ground states.  One of these is the
limit of the ground state as $\theta$ approaches $2\pi n$ from above;
the other is the limit as $\theta$ approaches $2\pi n$ from below.
So the charges of the two ground states are the two limiting values of
\juggo.

The analogous formula for right-moving fermions is easily derived, with
the result that for the same boundary conditions \fermibc\ the
right-moving fermion would contribute $-Q$.
Since the right-moving worldsheet fermions do carry non-vanishing left
$U(1)$ charge, it is important to take into account their contribution
when computing the left $U(1)$ charges of the twisted vacua.

We know the $U(1)$ charges $q$ of the fermions from Table 1, and in the sector
twisted by $g^{k}$ they pick up phases $\psi \rightarrow e^{-{i\pi kq}}\psi$
when going around the circle.  So without further ado, we can write the
general formula for the left $U(1)$ charges of the vacua:

\eqn\vaccharge{q_{k,-} =\sum_{i}\left\{ {(\alpha_{i}-1)}
\left({{k(\alpha_{i}-1)}\over 2}+
\left[{{k(1-\alpha_{i})}\over 2}\right] +{1\over 2}\right)
+{\alpha_{i}}~\left(-{k\alpha_{i} \over 2} + \left[
 {k\alpha_{i}\over 2}\right] + {1\over 2}\right)\right\} }

The analogous formula for the right-moving $U(1)$ charges is simply
\eqn\vacright{ {q_{k,+}} =
\sum_{i}\left\{ {\alpha_{i}}\left({{k(\alpha_{i}-1)}\over 2} +
\left[{{k(1-\alpha_{i})}\over 2}\right]+{1\over 2}\right)
+{{(\alpha_{i}-1)}}
\left(-{k\alpha_{i}\over 2} +\left[ {k\alpha_{i}\over 2}\right]
+{1\over 2}\right)\right\} }

We also need to determine the ground state eigenvalues of $L_{0-}$
($L_{0+}$ always vanishes in the ground state by right-moving supersymmetry).
In the (R,R) sectors, the vacuum eigenvalue of $L_{0-}$ vanishes.
Indeed, the
contribution of the fields in the ``internal'' Landau-Ginzburg theory
vanishes by supersymmetry, since the bosons and fermions satisfy the
same boundary conditions in (R,R) sectors.  The contribution of the 16
$E_{8}$ fermions (in their ground state, which is in the NS sector,
that is with antiperiodic boundary conditions)
is $-{16 \over 48}$ while the contribution
of the
10 $SO(10)$ fermions is ${10 \over 24}$ and the contribution of
the remaining 2 spacetime bosons (in light-cone gauge) is $-{2\over 24}$.
Simply doing the arithmetic, this sums to 0.

The (NS,R) sectors, on the other hand, can have negative vacuum
energies.  The 10 $SO(10)$ fermions, 16 $E_{8}$ fermions, and  2 spacetime
bosons contribute $-{5\over 8}$ to the vacuum energy.  The contribution
of the internal Landau-Ginzburg theory can be determined by using the
standard formulae for the energy of a twisted boson or fermion. The
contribution to the ground state energy (normal ordering constant of
$L_{0}$) for a complex fermion twisted by
$\theta$ ($-\pi
\le \theta \le \pi$) with respect to being antiperiodic
\eqn\febc{\psi \rightarrow e^{i(\pi
+ \theta)}\psi}
is given by
\eqn\fermtw{E_{\theta} ~=~-{1\over 24} ~+~ {1\over 8}~ ({\theta \over \pi})^{2}
{}~.}
A boson with the same boundary conditions would contribute the negative
of \fermtw ~to the vacuum energy.

We are interested in bose-fermi pairs with left $U(1)$ charges
$\alpha_{i}$ and $\alpha_{i} - 1$.
Therefore, if in some (NS,R) sector the fermion has boundary condition
$\psi \rightarrow e^{i(\pi + \theta )}\psi$
(with $\theta$ between $-\pi$ and $\pi$)
then the boson
is $\pi - |\theta |$ away from being antiperiodic.
Simply using the formula \fermtw ~we see that the fermion-boson pair then
contributes
\eqn\inten{E_{\theta } ~=~{1\over 4}{|\theta | \over \pi} - {1\over 8}}
to the vacuum energy.

Using these formulae and the fermion and boson $U(1)$ charges from table
1,
we find that
the vacuum energy of the sector twisted by $g^{k}$
with $k$ odd is given
in general by
\eqn\vacenergy{E_{k} ~=~-{5\over 8}~+~\sum_{i}
{}~\left({1\over 4}|{\beta^{i}_{k}}| -~ {1\over8}\right)~. }
$\beta^{i}_{k}$ is $k\alpha_{i}$, reduced mod $2$ to lie between $-1$ and
1.

Now that we know the quantum numbers of the twisted vacua $|0\rangle
_{k}$, we must determine the spectrum of physical states in each
twisted sector.  In the next section, we will do this in detail in a
familiar example: The Landau-Ginzburg model that corresponds to a quintic
hypersurface in ${\IC}\IP^4$.

\subsec{$E_6$ And Supersymmetry Multiplets And $U(1)$ Charges}

Certain symmetries of these systems -- $E_6$ symmetry and space-time
supersymmetry -- are not manifest in the formalism.  The proper
assembly of states into $E_6$ multiplets and supermultiplets can
be carried out using the $U(1)$ charges.

Let us consider first the construction of $E_6$ multiplets.
The $\bf{27}$ and $\bar {\bf{27}}$
of $E_{6}$ decompose under $SO(10)\times U(1)$ as
${\bf {27}} = {\bf {16_{1/2}\oplus 10_{-1} \oplus 1_{2}}}$ and
${\bf {\bar{27}}} = {\bf {\bar{16}_{-1/2}} \oplus 10_{1} \oplus 1_{-2}}$.
Therefore,
singlets of $SO(10)$ with $q_{-}=\pm 2$ are parts of $\bf{27}$s and
$\bar {\bf{27}}$s
of $E_{6}$, while singlets of $SO(10)$ with $q_{-}=0$ are
also singlets of $E_{6}$.
The decomposition of the adjoint representation of $E_6$
as $\bf{78}=\bf{45\oplus 16_{-3/2}\oplus \bar{16}_{3/2}\oplus 1}$ is also
helpful in studying gluinos.

The right-moving $U(1)$ charge plays a similar role in identifying
supermultiplets
\Lance .   For right-moving NS states, one can understand the
values of $q_{+}$ by considering unitarity constraints.  For example,
if we consider a state of right conformal weight $h_{+}$ and right-moving
$U(1)$
charge $q_{+}$, denoted by $|h_{+},q_{+}\rangle $, then using
\eqn\rsusycom{\eqalign{& \{ G_{1/2,+}, \bar G_{-1/2,+} \} ~=~2L_{0,+} + J_{0,R}
 \cr
& \{ G_{-1/2,+}, \bar G_{1/2,+} \} ~=~2L_{0,+} - J_{0,R} }}
and requiring that the states $G_{-1/2,+}|h_{+},q_{+}\rangle$ and
$\bar G_{-1/2,+}|h_{+},q_{+}\rangle$ have non-negative norm we find that
\eqn\const{h_{+} \geq {1\over 2}|q_{+}|~. }

This is useful because we know that massless right NS states must have
$h_{+}={1\over 2}$.  Then also requiring locality means that
$q_{+} = \pm 1$: if $q_{+}=1$, the state is right chiral
(annihilated by $\bar G_{-1/2,+}$) and if $q=-1$ the state is right
antichiral (annihilated by $G_{-1/2,+}$).

Consider a spin zero physical state $s$ of
$q_{+}=1$. It is represented by the spin zero part of a chiral
superfields $S$ with component expansion
\eqn\spacechi{S(x,\theta)  ~=~s(x) ~
+~\theta ~\eta (x) ~+~ \theta \theta ~ F(x)}
Likewise a scalar $\overline s$ of $q_+=-1$ is represented by
a supermultiplet
\eqn\santichi{\bar S(x,{\bar \theta)} ~=~\bar s(x)~+~\bar \theta ~ \bar \eta
(x)
{}~+~\bar \theta \bar \theta ~\bar F(x)~.}

We are most interested in the worldsheet quantum numbers of the
vertex operators for $\eta$ and $\bar \eta$, since we are going to be
finding the spectrum of spacetime fermions.  The fermions are obtained
by acting with the spacetime supersymmetries on \spacechi ~and
\santichi .  In particular, with the information derived above and a
knowledge of $U(1)$ charges of the spacetime supersymmetry generators, we
can infer the expected values of $q_{+}$ for the fermions which are part
of chiral or antichiral multiplets.
Recall that the explicit form of the spacetime supersymmetries is
\eqn\ssusy {{\eqalign{ &Q_{\alpha} ~=~\oint ~dz ~e^{-{\rho\over 2}}~S_{\alpha}
{}~\Sigma (z) \cr
& Q_{\dot\alpha} ~=~\oint~dz~e^{-{\rho\over 2}}~S_{\dot\alpha}~
\Sigma^{\dagger} (z) ~}}}
where $e^{-{\rho\over 2 }}$ is a spin field for the superconformal
ghosts, $S_{\alpha}$ and $S_{\dot \alpha}$ are spin fields for the world
sheet ``spacetime'' fermions $\psi^{\mu}$, and
$\Sigma$ and $\Sigma ^{\dagger}$
are Ramond sector fields which essentially implement right
spectral flow by $e^{\pm i\pi J_{0,R}}$.
Therefore, we see that $Q_{\alpha}$ and $Q_{\dot \alpha}$ leave the
value of $q_{-}$ unchanged, while they change $q_{+}$ by $\pm {3\over
2}$.

Now using the fact that $s$ is constrained to have $q_{+}=1$ by the
representation theory of the right moving N=2 algebra, we see that
$\eta$ must have $q_{+}=-{1\over 2}$, while the
vertex operator for the auxiliary field $F$ must have $q_{+}=-2$.
Similarly, $\bar \eta$ must have
$q_{+}={1\over 2}$, while $\bar F$ must have
$q_{+}=2$.

The same argument can be applied to find the quantum numbers of the
gauginos.  We know that generically
in heterotic string theory
the spacetime gauge symmetry must be generated
by (NS,NS) vector bosons, which correspond to states of the form
\eqn\vectboson{{\cal J}_{-1,L}~\psi^{\mu}_{-1/2,+}|0\rangle }
where ${\cal J}_{L}$ is a left-moving
symmetry generator and $\psi_{+}^{\mu}$ is one
of the right-moving ``spacetime''  fermions.
In particular, the state \vectboson ~always has $q_{+}=0$.  The
gauginos arise by applying the supersymmetries
\ssusy ~to the vector superfields, which have the same quantum numbers
as \vectboson .  Therefore, in particular gauginos always have
$q_{+} = \pm 3/2$.  For the gaugino partners of the $U(1)$ symmetries of
Gepner models, which are also neutral under the spacetime $E_6$ gauge
symmetry, $q_{-}=0$ as well.

So in summary: We expect to find fermions with $q_{+}~=~\pm {1\over
2}$ which are parts of spacetime antichiral and chiral supermultiplets,
and fermions with $q_{+}~=~\pm {3\over 2}$ which are part of spacetime
vector supermultiplets.  The latter are in correspondence with
generators of spacetime gauge symmetries.

\newsec {The Quintic}

Let us now use the technology developed in \S2 to study the
massless spectrum of string theory compactified on
a quintic hypersurface  $\IP_{4}(5) \subset {\IC}\IP^{4}$, in the
Landau-Ginzburg orbifold formulation.
We consider a quintic defined by the zeroes of a generic
quintic polynomial
\eqn\umbo{
W={1\over 5}\sum_{i_1\dots i_5}w_{i_1\dots i_5}\Phi^{i_1}\dots\Phi^{i_5}.}
In practice, that means that we consider a Landau-Ginzburg orbifold
with $W$ as superpotential.
The general results involve a reduction to a description involving
finite matrices.  When we want to make the results completely
explicit, we will consider the example of the Fermat quintic,
with
\eqn\fourone{ W= \sum_{i=1}^{5} {1\over 5}~\Phi_{i}^{5}~
 }
which has enhanced symmetry and corresponds to a soluble Gepner point
\Gepner .
We will carry out the discussion for a $(2,2)$ model with superpotential
$W$, but no essential modification is required for the $(0,2)$ case,
as we will explain in \S3.9.

We must obtain the spectrum in 10 sectors, which arise,
starting with the untwisted (R,R) sector,
by twisting by $\exp(-ik\pi J_{0L})$, with $0\leq k\leq 9$.
In practice, it suffices to consider $0\leq k \leq 5$, as
CPT exchanges $k$ with $10-k$.

The (R,R) sector is the sum of the twisted sectors of even $k$,
and the (NS,R) sector is the sum of the twisted sectors of odd $k$.
Happily, the (R,NS) and (NS,NS) sectors need not be studied explicitly,
as they are related to (R,R) and (NS,R) by space-time supersymmetry.

As a preliminary, let us review the fields and their quantum numbers
here.  In addition to the bosons $\phi_{i}$ and $\bar \phi_{i}$, there
are left moving fermions $\psi^{i}_{-}$ and $\bar \psi^{i}_{-}$, and
right moving fermions $\psi^{i}_{+}$ and $\bar \psi^{i}_{+}$. Their left
and right moving $U(1)$ charges are summarized in Table 2:

$$\vbox {\settabs 3 \columns
\+& ${\bf Table ~2}$&\cr
\+&&\cr
\+{$\bf Field$} &{$\bf q_{-}$}&{$\bf q_{+}$}\cr
\+~~{${\phi_{i}}$}&${1/5}$&${1/5}$ \cr
\+~~{$\bar \phi_{i}$}&$-{1/5}$&$-{1/5}$ \cr
\+~~$\psi^{i}_{-}$ &$-{4/5}$&${1/5}$ \cr
\+~~$\psi^{i}_{+}$ &${1/5}$&$-{4/5}$ \cr
\+~~$\bar \psi^{i}_{-}$ &${4/5}$&$-{1/5}$ \cr
\+~~$\bar \psi^{i}_{+}$ &$-{1/5}$&${4/5}$ \cr}$$
\bigskip
So, using the general formula for $U(1)$ charges of ground states
developed in the last section, we find that the left and right $U(1)$
charges $q_{k,-}$ and ${q_{k,+}}$ of the twisted sector vacua are
\eqn\fourthree {q_{k,-}~=~5~\left\{
 -{4\over 5}~\left(-{2k\over 5} + \left[{2k\over 5}\right] + {1\over 2}\right)
{}~+~{1\over 5}~\left(-{k\over 10} + \left[{k\over 10}\right] + {1\over
2}\right)
 \right\} }
\eqn\qright{ {q_{k,+}} ~=~5~\left\{{1\over 5}~\left(-{2k\over 5}
+ \left[{2k\over 5}\right] + {1\over 2}\right)
{}~-~{4\over 5}~\left(-{k\over 10} + \left[ {k\over 10}\right] + {1\over
2}\right) \right\} }
where the fractional fermion numbers in \fourthree ~and \qright
{}~arise
because of the boundary conditions on the fermions in the sector twisted
by $g^{k}$
\eqn\fourfour{\psi^{i}_{-} \rightarrow e^{i{{4\pi}\over 5} k} \psi^{i}_{-}, ~
\psi^{i}_{+} \rightarrow e^{-i{\pi\over 5}k} \psi^{i}_{+}~.}

We also need to know the ground state energies of the vacua in the
various twisted sectors.  Using the normal formulae for the zero-point
energies of twisted bosons and fermions as discussed in \S2 , we
see that the even $k$ sectors have vanishing vacuum energy, as expected
from supersymmetry,
while the odd $k$ sectors have ground state energies
\eqn\fourfive{E_{1}=-1,~E_{3}=-{1\over 2},~E_{5}=0,~E_{7}
=~-{1\over 2},~E_{9}=-1 ~.}

We recall that upon
taking the $\bar Q_{+,R}$ cohomology, the right-moving fermions
are eliminated, so in this analysis the left-moving fermions are
the only ones of interest and will be usually
denoted as $\psi^i$, not $\psi^i_-$.
Also, upon taking the
$\bar Q_{+,R}$ cohomology, the zero modes of $\bar\phi$ are eliminated.
So in practice, we need to compute in the sectors $0\leq k\leq 5$
and in the reduced Hilbert space the cohomology of
the $\bar Q_{+,L}$ operator \foot{In our analysis of models, we
will drop the $i{\sqrt 2}$ prefactor of $\bq$ in \screening, which is
obviously irrelevant in computing the cohomology.}
\eqn\fourtwo{ \bar Q_{+,L} ~=~\sum_{i=1}^{5}~\oint
{}~{\partial W \over \partial \phi_{i}}\psi^{i}_{-}.}
After carrying out the analysis, we will summarize
the resulting spectrum at the end of this section.  We also will
assemble the
$\bf{1}$'s,~$\bf{10}$'s and $\bf{16}$'s of $SO(10)$ into $\bf{27}$'s
and $\bar {\bf{27}}$'s of $E_{6}$, using
the values of the left $U(1)$ charge as explained in \S2.5.

\subsec{$k=0$ Sector}

This corresponds to the normal untwisted (R,R) sector.  The ground state
energy vanishes.
Since all of the fields are untwisted, the relevant
lowest energy modes are (from the comment in the last paragraph)
the zero modes
\eqn\gurff{\phi^{i}_{0},~\psi^i_{0}, ~\bar \psi^{i}_{0}.}

The commutation relations of the fermion zero modes are
\eqn\urff{\{\psi^i_0,\bar\psi^j_0\}=\delta^{ij}.}
We let $|0\rangle$ denote a Fock vacuum with
\eqn\foursix{\psi^{i}_{0} |0\rangle ~=~0 ~.}
This state has left and right moving $U(1)$ charges $(q_{-},q_{+})=
(-3/2,-3/2)$.

Since the ground state energy is zero, in studying zero energy
states we can altogether ignore the oscillator modes in the definition
of $\bar Q_{+,L}$, so that $\bar Q_{+,L}$ reduces
to
\eqn\grog{\bar Q_{+,L}=\psi_0^i{\partial W(\phi_0)\over\partial\phi_0^i}.}
The cohomology of $\bar Q_{+,L}$ is generated entirely
by states of the form
\eqn\fourseven{F(\phi_{0}) ~|0\rangle}
and the projection onto half-integral $U(1)$ charges
means that we need consider
only functions $F$ of degree $5 j$ for $j = 0,1,2,\dots $.  But also, note that
\eqn\foureight{{\bar Q_{+,L}}~{\bar \psi^{i}_{0}} |0\rangle ~=~ {\partial
W\over \partial \phi_{i}}~|0\rangle }
so we must mod out by the ideal generated by the $\{ {\partial W\over
\partial \phi_{i}} \}$.
What we have found here is of course just the famous result that the
chiral ring $\cal R$ of a Landau-Ginzburg theory is given by the
``singularity ring'' of the superpotential
\eqn\fournine{{\cal R} ~\simeq ~{{\bf C}[\phi^i]
\over \{\partial_{j}W(\phi)
\}}~.}

It is easy to enumerate the resulting states.
At $(q_{-},q_{+})=(-3/2,-3/2)$, we simply get $|0\rangle$.
At $(-1/2,-1/2)$, we get the quintic functions of $\phi$ modulo the ideal
generated by derivatives of $W(\Phi)$ -- 101 states in all according to
a standard counting.
At $(1/2,1/2)$, we get the tenth order polynomials modulo those in the
ideal generated by the derivatives -- again 101 states.
At $(3/2,3/2)$, there is a single state; for instance, for the Fermat
polynomial, it can be represented by
$\prod_{i=1}^{5}~\phi_{i}^{3}|0\rangle$.

Making the GSO projections as described in \S2.3,
the states in this sector with $(q_{-},q_{+}) = (-1/2,-1/2)$ correspond
to right handed fermions in the $\overline{\bf{16}}$ of
$SO(10)$, while those with
$(q_{-},q_{+}) = (1/2, 1/2)$ correspond to left handed fermions in the
$\bf{16}$ of $SO(10)$.  In fact, the
former are the $\overline{\bf{16}}$ components of
the 101 right-handed $\bar {\bf{27}}$'s, while
the latter are the $\bf{16}$
components of the 101 left handed $\bf{27}$'s.
The $(-3/2,-3/2)$ and $(3/2,3/2)$
states are gluinos, according to the discussion of the right-moving
$U(1)$ charge in \S2.5. The GSO projections cause the $(-3/2,-3/2)$ states
to be left-handed in space-time and a $\bf{16}$ of $SO(10)$,
while the $(3/2,3/2)$
are a right-handed $\bar{\bf{16}}$.

\subsec {k=1 Sector}

The ground state energy is $E_{1}=-1$ and the ground state $U(1)$
charges are $(0,-3/2)$.
Because of the twist by $e^{-i\pi J_{0}}$, there are no
zero modes.  However, since we are looking for states of energy 0
and the vacuum has negative energy, we should keep the lowest excited
modes of the fields.  These are
\eqn\fourten{\phi^{i}_{-1/10},~\bar \phi^{i}_{-9/10},~\psi^{i}_{-3/5},~\bar
\psi^{i}_{-2/5}~.}

Now, we simply need to write down all the states that have zero energy
in the free field approximation
and find the $\bar Q_{+,L}$ cohomology.
The $\bar Q_{+,L}$ operator restricted to the relevant modes is
\eqn\roar{\eqalign{\bar Q_{+,L}= &
\psi^{i_{1}}_{2/5}w_{i_{1}\dots_{i_{5}}}\phi^
 {i_{2}}_{-1/10} \dots \phi^{i_{5}}_{-1/10}\cr
&+ 4\psi^{i_{1}}_{-3/5}w_{i_{1}\dots i_{5}}\phi^{i_{2}}_{9/10}
\phi^{i_{3}}_{-1/10}\phi^{i_{4}}_{-1/10}
\phi^{i_{5}}_{-1/10} ~.}}
Other terms in $\bar Q_{+,L}$ have zero matrix elements among states
of zero energy.
Since $\bar Q_{+,L}$ does not
change left $U(1)$ charge, we can compute its cohomology separately in
spaces of states of different $q_{-}$.

First we consider states constructed without $\lambda$ excitations.
These will be $SO(10)$ singlets; they either have $q_{-}=2$
and are part of $\bf{27}$'s  of $E_6$, or
$q_{-}=0$ and are $E_6$ singlets.

\vglue 10pt
\noindent\underbar{\raise 3pt\hbox{$SO(10)$ Singlet Components Of
$\bf{27}$'s
}}
\vglue 2pt

There are three types of zero energy states at $q_{-}=2$:
\eqn\exhau{\eqalign{
&{q_{+}}=-3/2: ~\phi^{i}_{-1/10}\phi^{j}_{-1/10}\bar \psi^{k}_{-2/5}
\bar \psi^{l}_{-2/5}|0\rangle
{}~~~~~~ (150)\cr &{q_{+}}=-1/2: ~\phi^{i_{1}}_{-1/10}\dots\phi^{i_{6}}_{-1/10
 }\bar\psi^{j}_{-2/5}|0\rangle
{}~~~~~~(1050) \cr
&{q_{+}}=1/2: ~\phi^{i_{1}}_{-1/10}\dots\phi^{i_{10}}_{-1/10}|0\rangle
{}~~~~~~(1001) }}
where the numbers represent the number of distinct states of each
type.
$\bar Q_{+,L}$ increases the
value of $q_{+}$ by one, so we have a sequence of
maps
\eqn\sequence {0 ~\rightarrow ~V_{-3/2} ~^{\bar Q_{+,L} \atop \rightarrow}
{}~ V_{-1/2}~^{\bar Q_{+,L}
\atop \rightarrow}~V_{1/2}~\rightarrow ~0~.}
Here, $V_{q_{+}}$ denotes the space spanned by the states of right $U(1)$
charge $q_{+}$.

In the general case, one can write down a similar sequence to \sequence ~above.
For each fixed value of the left $U(1)$ charge,  one gets a
sequence
\eqn\genseq {0 ~\rightarrow V_{q_{+}}~^{\bar Q_{+,L} \atop \rightarrow}
{}~V_{{q_{+}}+1}~^{\bar Q_{+,L}\atop\rightarrow} ~\dots ~^{\bar Q_{+,L}
\atop \rightarrow}~V_{{q_{+}}+n} ~\rightarrow ~0}
where in general $V_{q_{+}}$ is the space of
states with right $U(1)$ charge ${
{q_{+}}}$.
Then the $\bar Q_+$ cohomology is the cohomology of \genseq.

The concrete case of \sequence\ can be analyzed as follows.
The states of $q_{+}=-3/2$ can be written in the form
${\bar \psi_{-2/5,i}\bar\psi_{-2/5,j}A^{ij}(\phi_{-1/10})}|0\rangle$
where $A^{ij}(\phi_{-1/10})$ are homogeneous quadratic functions of
$\phi_{-1/10}$, in components $A^{ij}=A^{ij}{}_{kl}\phi^k_{-1/10}
\phi^l_{-1/10}$.
The states of $q_{+}=-1/2$ are of the form
$\bar\phi_{-2/5,i}B^i(\phi_{-1/10})|0\rangle$,
with the $B_i$ being homogeneous sixth order functions, and  the states
of $q_{+}=1/2$ are of the form $C(\phi_{-1/10})|0\rangle$, with
$C$ being a homogeneous tenth order function.  The action of the $\bar Q_+$
operator is
\eqn\actop{\eqalign{
\bar Q_{+,L}\left(\bar\psi_{-2/5,i}\bar\psi_{-2/5,j}A^{ij}|0\rangle\right) &
= \bar \psi_{-2/5,j}A^{ij}{\partial W\over\partial\phi_i}|0\rangle \cr
\bar Q_{+,L}\left(\bar\psi_{-2/5,i}B^i |0\rangle
\right) & = B^i{\partial W\over\partial
\phi^i}|0\rangle.\cr}}
This is precisely isomorphic to a piece of the effective $\bar Q_{+,L}$
operator that we met in the $k=0$ sector, except that the variables are
now called $\phi_{-1/10}$ instead of $\phi_0$ and $\bar\psi_{-2/5}$
instead of $\bar\psi_0$.  In particular, the cohomology vanishes
at $q_{+}=-3/2$ and $-1/2$, and at $q_{+} =1/2$, the cohomology
consists of the tenth order polynomials in $\phi_{-1/10}$ modulo
the ideal generated by $\partial_i W$.  This is a 101
dimensional space, in natural one-to-one correspondence with the Ramond
ground states of $k=0$ that were constructed from tenth order polynomials.
This is expected from $E_6$ symmetry: these states will combine with
some of the $k=0$ states into $E_6$ multiplets.

\vglue 10pt
\noindent\underbar{\raise 3pt\hbox{$E_6$ Singlets
}}
\vglue 2pt

$E_6$ singlets arise as $SO(10)$ singlets of $q=0$.
These take the following form:
\eqn\singstates{\eqalign{ &{q_{+}}=-3/2:~\bar \psi_{-2/5,i}\psi^{j}_{-3/5}
|0\rangle~~~~~~(25)~\cr &~~~~~~~~
 {\rm and~also}~~\bar \phi_{-9/10,j}  \phi^{i}_{-1/10}|0\rangle
{}~~~~~~(25) \cr
&{ q_{+}}=-1/2:
{}~\phi^{i_{1}}_{-1/10}\dots\phi^{i_{4}}_{-1/10}\psi^{j}_{-3/5}|0\rangle
{}~~~~~~~(350)~.}}
The number in parentheses is the number of states of a given type.

The maps in the resulting sequence of the form \genseq ~are given by
\eqn\moremaps{\eqalign{&  \bq \left(A_i{}^j \bar \psi_{-2/5,j}\psi^{i}_{-3/5}
|0\rangle
\right) = A_i{}^j\psi^{i}_{-3/5} \partial_jW(\phi_{-1/10})|0\rangle\cr}}
\eqn\more{\eqalign{
&\bq \left(B_i{}^j \bar \phi_{-9/10,j}\phi^{i}_{-1/10}|0\rangle \right) =
{}~~~~~ B_i{}^j
\phi^i_{-1/10}\psi^k_{-3/5}\partial_j\partial_kW(\phi_{-1/10})
|0\rangle .\cr}}

In the particular case of the Fermat quintic \fourone , one can see that
at $q_+=-3/2$, $\bar Q_+$
has a five dimensional kernel,
spanned by the states $({1\over 4}\phi^{i}_{-1/10}\bar\phi_{-9/10,i} -
\bar \psi_{-2/5,i}\psi^{i}_{-3/5})|0\rangle$ (no sum over $i$).
This means that 45 of
the states at $q_{+}=-1/2$ are trivial in $\bq$ cohomology
while the other 305 must represent nontrivial cohomology classes.
Hence, one finds 310 singlets of $SO(10)$ in this
sector for the Fermat quintic: 5 at $q_+=-3/2$ and 305 at $q_+=-1/2$.
One of the states at $q_+=-3/2$ is present for generic $W$ and is
in the adjoint representation of $E_6$; the other 4 at $q_+=-3/2$
and all 305 at $q_+=-1/2$ are singlets of $E_6$.

In general, to summarize equations \moremaps,\more,
the $E_6$ singlets at $q_{+}=-1/2$ are represented by
a collection of five quartic functions $P_i(\phi_{-1/10})$ subject
to the equivalence relation
\eqn\mormo{P_i\cong P_i +A_i{}^j{\partial W\over\partial \phi^j}+\phi^kB_{kl}
{\partial^2W\over\partial\phi^l\partial\phi^i}.}
After redefining $A$, this can alternatively be written
\eqn\ormo{P_i\cong P_i+A_i{}^j\partial_jW+\partial_i\left(\phi^k B_k{}^l
\partial_l W\right).}

\vglue 10pt
\noindent\underbar{\raise 3pt\hbox{Finer Classification Of $E_6$
Singlets
}}
\vglue 2pt

In the field theory limit, there are three types of massless
$E_6$ singlets at $q_{+}=-1/2$, namely states that originate in $H^1(\QP,T)$,
$H^1(\QP,T^*)$, and $H^1(\QP,{\rm End}(T))$.
These may be distinguished as follows.
States $|\Psi\rangle$ which satisfy the chiral condition
$$G_{-1/2}|\Psi\rangle = 0$$
correspond to elements of $H^{1}(\QP,T)$ while those which satisfy
the anti-chiral condition
$$\bar G_{-1/2}|\Psi\rangle = 0$$
correspond to elements of $H^{1}(\QP,T^{*})$.
The singlets which are orthogonal to those obeying the chiral
or anti-chiral condition correspond to elements of $H^{1}(\QP, {\rm End
(T)})$.  We want to implement this classification in the Landau-Ginzburg
model, using the explicit forms of $G_{-1/2}$ and $\bar G_{-1/2}$ from
\omico.

First of all, using the above explicit description of the $E_6$ singlets
at $q_{+}=-1/2$, and the fact that $\bar G_{-1/2}$ has a term
proportional to $\phi^i_{-1/10}\bar\psi_{-2/5,i}$, none of the
$E_6$ singlets of $q_{+}=-1/2$ are annihilated by $\bar G_{-1/2}$.
On the other hand, one finds that
\eqn\morefin{G_{-1/2}\left(\psi^i_{-3/5}P_i(\phi_{-1/10})|0\rangle\right)
\sim \psi^i_{-3/5}\psi^j_{-3/5}\partial_iP_j(\phi_{-1/10})|0\rangle.
 }
This therefore vanishes precisely if $\partial_iP_j-\partial_jP_i=0$,
or in other words if $P_i=\partial_iS$ for some quintic polynomial
$S$. From the homogeneity of the $P_i$ it follows that
$S=\phi^iP_i/5$.  The equivalence relation \mormo\ then amounts
to
\eqn\tormo{S\cong S+ \phi^iA_i{}^j\partial_jW +\phi^i\phi^kB_{k}{}^l
{\partial^2W\over\partial\phi^l\partial\phi^i}.}
{}From the homogeneity of $W$ it follows that
$\phi^i\partial_i\partial_lW
=4\partial_lW$, and finally then the space of $E_6$ singlets annihilated
by $G_{-1/2}$ is the 101 dimensional space of quintic polynomials $S$
modulo the usual ideal generated by the $\partial_iW$.

Now we want to look at the analog of $H^1(\QP,{\rm End}(T))$ -- the states
orthogonal to the chiral and anti-chiral states.  Since we have
already taken account of the states of the form $\partial_iS$, we now
look at states $P_i$ with anything of the form $\partial_iS$ considered
trivial.  Hence the analog of $H^1(\QP,{\rm End}(T))$ in the $k=1$ sector
is the space of five quartic polynomials $P_i$ subject to
\eqn\unno{P_i  \to P_i+A_i{}^j\partial_jW+\partial_iS.}
(This is similar to \ormo\ but $\phi^iC_i{}^j\partial_kW$ is now
replaced by an arbitrary quintic $S$.)
By homogeneity of $S$, $\phi^i\partial_iS=5S$, so $S$ can be uniquely
fixed by normalizing $P$ so that $\phi^iP_i=0$.
So the analog of $H^1(\QP,{\rm End}(T))$ can be identified with the
space of five quartic polynomials $P_i$, with $\phi^iP_i=0$, and the
equivalence relation
\eqn\uppo{P_i\cong P_i +A_i{}^j\partial_jW-{1\over 5}\partial_i\left(\phi^k
A_k{}^j\partial_jW\right).}

Now let us compare this to the computation of
$H^1(\QP,{\rm End}(T))$ in the field theory limit.
A tangent vector to the quintic hypersurface $W=0$ in $\IC\IP^4$
can be represented by a collection of five complex numbers
$V^i$ obeying an equivalence relation
\eqn\eqrln{V^i\to V^i+\lambda \phi^i }
(with $\phi^i$ being the homogeneous coordinates on $\IC\IP^4$)
and a constraint
\eqn\lolly{V^i\partial_iW = 0 }
(so that the vector field on $\IC\IP^4$ represented by the $V^i$ is
tangent to the hypersurface $W=0$).  The constraint \lolly\
and equivalence relation \eqrln\ are compatible because
$\phi^i\partial_iW=5W$ vanishes at $W=0$.  To deform the tangent
bundle of the quintic, one can replace
\lolly\ by
\eqn\tolly{V^i(\partial_iW+P_i)=0}
where the $P_i$ are homogeneous quartic polynomials and
(to maintain compatibility with \eqrln) $\phi^iP_i=0$.
So in field theory, $H^1(\QP,{\rm End}(T))$ is the space of $P_i$'s
subject to $\phi^iP_i=0$.  The answer is almost the same in the
Landau-Ginzburg model, but in the Landau-Ginzburg theory
there is an additional equivalence relation
\uppo, so some states are missing.  We will return to this point
after examining the spectrum for other values of $k$.

For the time being, let us just quantify the discrepancy.
For generic $W$, the equation
\eqn\ilo{A_i{}^j\partial_jW-{1\over 5}\partial_i\left(\phi^k
A_k{}^j\partial_jW\right)=0}
is obeyed only if $A_i{}^j=\delta_i{}^j$.  In that case, the Landau-Ginzburg
theory is missing 24 states with the quantum numbers of a
traceless matrix $A_i{}^j$.  It can happen that for particular
$W$'s there are other $A$'s for which \ilo\ vanishes.  In that case
the equivalence relation
\uppo\ is less powerful, so the Landau-Ginzburg theory has extra
massless $E_6$
singlets (for example,
we saw that in the case of the Fermat quintic there are five extra states
$({1\over 4}\phi^{i}_{-1/10}\bar\phi_{-9/10,i} -
\bar\psi_{-2/5,i}\psi^{i}_{-3/5})|0\rangle$ in the $\bq$
cohomology).  When this happens, there are extra massless $E_6$ singlets
at $q_+=-3/2$ that are supersymmetric partners of extra gauge bosons
that occur for this particular $W$, and extra massless singlets
at $q_+=-1/2$ that are supersymmetric partners of Higgs bosons that
will give mass to the extra gauge bosons when $W$ is perturbed.
Apart from this possibility of extra gauge symmetries and scalar partners
for particular
$W$'s, the discrepancy between field theory and $k=1$ Landau-Ginzburg
theory consists of 24 missing states with the quantum numbers of a traceless
matrix $A_i{}^j$.

\vglue 10pt
\noindent\underbar{\raise 3pt\hbox{$SO(10)$ $\bf{10}$ Components Of
$\bar {\bf{27}}$'s
}}
\vglue 2pt

The states we have been considering so far have all been $\bf{1}$'s of
$SO(10)$,
but we also need to consider $\bf{10}$'s of $SO(10)$.  Such states will contain
 an
excitation of the gauge fermions $\lambda ^{I}_{-1/2}$
and will correspond to cohomology classes of $\bar Q_{+,L}$ with
total energy $-1/2$ in the internal theory.  We find two patterns of
such
states, both with $q_{-}=1$:
\eqn\tens{\eqalign { &{q_{+}}=-3/2 :~
\lambda_{-1/2,I}\cdot\phi^{i}_{-1/10}\bar
\psi_{-2/5,j}|0\rangle ~~~~~~~(25) \cr
&{q_{+}}=-1/2:~\lambda_{-1/2,I}
\phi^{i_{1}}_{-1/10}\dots\phi^{i_{5}}_{-1/10}|0\rangle
{}~~~~~(126) ~.}}
The states of $q_{+}=-1/2$ can thus be written as $\lambda_{-1/2,I}
S(\phi_{-1/10})|0\rangle$ with $S$ a homogeneous quintic function.
The map in the associated sequence \genseq ~is given by
\eqn\tenmap {\eqalign { &\bq \left( \lambda_{-1/2,I}C_i{}^j\phi^{i}_{-1/10}\bar
\psi_{-2/5,j}\right)|0\rangle=\lambda_{-1/2,I}\phi^i_{-1/10}C_i{}^j
\partial_jW(\phi_{-1/10})|0\rangle
}}
The cohomology is thus the space of quintic homogeneous polynomials
$S(\phi_{-1/10})$ modulo the ideal generated by the $\partial_jW$.
This is the familiar space of Ramond ground states at $k=0$ -- to which
these are indeed related by $E_6$ symmetry.
In fact, these $\bf{10}$'s of $SO(10)$ have a simple (and standard) relation
to the $E_6$ singlets with $P_i=\partial_iS$
that are annihilated by $G_{-1/2}$ and derived from $H^1(\QP,T)$ in the field
theory limit.  The $E_6$ singlets arise by acting on $S(\phi_{-1/10})|0\rangle$
with $G_{-1/2}$, and the $\bf{10}$'s of $SO(10)$ arise by acting on the same
states with $\lambda_{-1/2}$.

\vglue 10pt
\noindent\underbar{\raise 3pt\hbox { Other States
}}
\vglue 2pt

The states $\lambda_{-1/2,i}\lambda_{-1/2,j}|0\rangle$
have $U(1)$ charges $(0,-3/2)$ and are left-handed gluinos
in the adjoint representation of $SO(10)$.

The states $\partial_{-}X^{\mu}_{-1}|0\rangle$, where $X^{\mu}$ are the
Minkowski space bosons, represent the left-handed gravitino and
dilatino.

Gluinos of the second $E_8$ have the form $\tilde J^a_{-1}|0\rangle$,
where $\tilde J^a$ are the left-moving world-sheet currents generating
the second $ E_8$.

This completes the analysis of the massless fermions for $k=1$.

\subsec{$k=2$ Sector}

This sector has vanishing ground state energy and $(q_{-},q_{+}) =
(3/2,-3/2)$  as the ground
state $U(1)$ charges.  All of the fields are twisted, so there are no zero
modes and hence we get only one state of total energy zero, the ground
state.
This single element of $\bq$ cohomology from the $k=2$ sector is a left
handed $\overline{\bf{16}}$ of $SO(10)$; these
are gluinos forming part of
the adjoint representation of $E_6$.

\subsec{$k=3$ Sector}

The ground state energy is $-1/2$ and the ground state $U(1)$ charges are
$(q_{-},q_{+})=(-1,-1/2)$.
The lowest modes of the various fields are
$$\phi^{i}_{-3/10},~\bar \phi_{-7/10,i},~\psi^{i}_{-4/5},~\bar
\psi^{i}_{-1/5}~.$$
These values ensure the important fact that $G_{-1/2}$ annihilates
the ground state in this sector.  However,
\eqn\umpo{\bar G_{-1/2}|0\rangle \sim \sum_{i=1}^5\bar\psi_{-1/5,i}\,\,
\,\phi^i_{-3/10}|0\rangle}
does not vanish.  As it has zero energy and $q_{-}=0$, and is obviously
annihilated by $\bar G_{-1/2}$, it is an $E_6$ singlet related to
$H^1(\QP,T^*)$.

The only other states of vanishing energy built out of ``internal''
excitations are the $(q_{-},q_{+})=(0,-1/2)$ states
$$A_j{}^i\bar \psi_{-1/5,i}\phi^{j}_{-3/10}|0\rangle $$
with a traceless matrix $A_i{}^j$.  These are annihilated by
neither $G_{-1/2}$ nor $\bar G_{-1/2}$
so they are analogous to $H^1(\QP,{\rm End}(T))$ in field theory.
Indeed, we have found the piece of $H^1(\QP, {\rm End}(T))$ that
was missing in the $k=1$ sector.

Actually, because of instanton effects,
a precise correspondence between the classical $H^1(\QP,{\rm End}(T))$
and the Landau-Ginzburg contribution was not guaranteed and does not
occur in general; we do not know why it occurs in the particular
case of the quintic hypersurface.  However, one is guaranteed that the
``character-valued'' index (the imaginary part of the character of any
discrete symmetries that may be present in field theory, for a particular
$W$) should be the same for field theory or Landau-Ginzburg, since this
index is a topological invariant.\foot{More
generally, the element in the {\bf K} theory of the moduli space
of complex structures represented by the left-handed singlets minus the
right-handed ones is a topological invariant.}
The missing piece that we have just
found was the simplest possibility compatible with this topological
invariance.

We can also act with the gauge fermions $\lambda_{-1/2,I}$ on the
vacuum $|0\rangle$ to produce a single $\bf{10}$ of $SO(10)$ which is also
in the cohomology of $\bar
Q_{+,L}$. Since this state
has $q_{+}=-1/2$, it corresponds to a right handed fermion.
Of course, this state has the usual relation to the anti-chiral
state \umpo; one is obtained by acting on a suitable state
(here the vacuum) by $\bar G_{-1/2}$, while the other is obtained
by acting on the same state with $\lambda_{-1/2,I}$.

\subsec{$k=4$ Sector}

The ground state has zero energy and $(q_{-},q_{+})=(1/2,-1/2)$.  Since all
fields are twisted, there are no zero modes and the ground state is the
only state of zero energy we can construct.  So this sector contributes
to the $\bar Q_{+,L}$ cohomology one right handed $\bf{16}$ of $SO(10)$, which
is part of a $\bf{27}$ of $E_{6}$.

\subsec{$k=5$ Sector}

The ground state has zero energy again.
Actually, the ground state is not unique because
$\psi_-$ and $\bar \psi_-$ are untwisted and have zero modes.
The state $|0\rangle$ annihilated by $\bar \psi_{0,i}$ has $U(1)$
charges $(2,-1/2)$.  Other states are obtained by acting with
factors $\psi^i_0$.  As this field has quantum numbers $(-4/5,1/5)$,
the only state other than $|0\rangle$ that has integral $q_{-}$ and
hence survives in the orbifold is the state $\prod_{i=1}^5
\psi^i_0|0\rangle$, with quantum numbers $(-2,1/2)$.
So this sector contributes two states, both singlets of $SO(10)$,
one part of a
right handed $\bf{27}$ and one part of a left handed ${\bar{\bf{27}}}$
of $E_6$.

\subsec{Summary: Spectrum Of String Theory On $M^{4}\times \IP_{4}(5)$}

The rest of the massless spectrum (for $k>5$)
follows by complex conjugation
from the above results.

Assembling the pieces, for the Fermat superpotential
there are 330
$E_{6}$ singlets that are superpartners of massless scalars (1 coming
from $H^{1}(\QP,T^*)$, 101 coming from $H^1(\QP,T)$, and 228 coming from
$H^1(\QP,{\rm End}(T))$), 4 that
are superpartners of neutral gluinos,
101 left handed ${\bf{27}}$'s of $E_{6}$, and 1
left handed $\bar {\bf{27}}$ of $E_{6}$ (along with their right handed
anti-particles).  Those
numbers agree with those found by Gepner in his analysis of this
model as the product of five level three minimal models.
In particular, the enhanced gauge symmetry ($U(1)^4$ associated with the
four neutral gluinos) agrees with that found by Gepner.

One can see the $SO(10)$ multiplets combine into multiplets of $E_{6}$
more explicitly, as follows.
Under $SO(10)\times U(1)$, the $\bar {\bf{27}}$'s
decompose as ${\bf{\overline {1
 6}}_{-1/2}
\oplus 10_{1} \oplus 1_{-2}}$.
The way the $\bar {\bf{27}}$'s arise in this model is indicated in Table 3:

$$\vbox {\settabs 4 \columns
\+&{\hfill \bf Table 3}&& \cr
\+&&&\cr
\+Sector&$\bf {1_{-2}}$&$\bf {\overline {16}_{-1/2}}$&$\bf {10_{1}}$ \cr
\+~~~0&0&101&0 \cr
\+~~~1&0&0&101\cr
\+~~~2&0&0&0 \cr
\+~~~3&0&0&0 \cr
\+~~~4&0&0&0 \cr
\+~~~5&1&0&0 \cr
\+~~~6&0&1&0 \cr
\+~~~7&0&0&1 \cr
\+~~~8&0&0&0 \cr
\+~~~9&101&0&0 \cr}
$$
\bigskip
The table shows the number of $SO(10)\times U(1)$ multiplets of given
type arising in each sector.
In general, starting
from the $\bf{10}$, one obtains the
$\overline{\bf{16}}$ by spectral flow
by $e^{i\pi J_{0}}$ and the $\bf{1}$ by spectral flow by
another $e^{i\pi J_{0}}$; the sector number $k$ shifts by 1 each time.
One important point is not indicated in the above table:
The $\bar {\bf{27}}$'s
coming from sectors 0, 1, and 9 are right-handed in space-time while
the $\bar {\bf{27}}$ coming from sectors 5, 6, and 7 is left-handed.
The corresponding table for $\bf{27}$'s comes by complex conjugation,
and the analogous table for gluinos can be similarly constructed.

\subsec{Absence Of Anomalies In The $\IZ_{5}$ Symmetry}

Part of the fascination of the Landau-Ginzburg models is that they
have a ``quantum'' symmetry, not
present for other choices of the Kahler class,
which keeps track of the sector number $k$.
This ${\IZ}_{10}$ symmetry
is the product of a ${\IZ}_2$ symmetry (which counts
fermion number modulo two and is always present) and
a quantum ${\IZ}_5$ symmetry.  It can be seen that this symmetry
is actually an $R$ symmetry in space-time.

A natural question is whether the quantum symmetry suffers from an anomaly
at  the level of space-time instantons.  To answer this question,
it suffices to consider only instantons contained inside $SO(10)$.
In units in which a left-handed fermion multiplet in the ${\bf 10}$
of $SO(10)$ contributes 1 to the anomaly, the ${\bf 16}$ and ${\bf\overline
{16}}$ contribute 2 and the ${\bf 45}$ contributes 8.  Working
out the values of $k$ for the various left-handed multiplets (and remembering
to include the gluinos), one
finds that the quantum symmetry has no anomaly for $E_6$ instantons
(and also no anomaly for instantons in the second $E_8$).

\subsec {(0,2) Deformations}

$(0,2)$ deformations of the quintic can be constructed by deforming the
tangent bundle as a holomorphic vector bundle over $X$.  As we have
recalled in \tolly, this is done by substituting $\partial_iW\to
\partial_iW + G_i$ (where $G_i$ are quartic polynomials obeying
$\phi^iG_i=0$) in the definition of the tangent bundle.
As one can see from \Witten, \S6, the effect of this on the $\bar Q_{+,L}$
operator will be just the obvious substitution; the $\bar Q_{+,L}$ operator
of the $(0,2)$ model is simply
\eqn\screentwo {\bar Q_{+,L} ~=~i\sqrt 2
\oint~\sum_{i}~\left({\partial W\over \partial
\phi_{i}}+G_i\right)\psi^{i}_{-}.}
Our techniques then carry over to the (0,2) case without any conceptual
difficulties.
The physical spectrum of
the (0,2) model is given by the cohomology of $\bar Q_{+,L}$, which can
be computed by the same methods that we have used at $G=0$.

\newsec {Directions For Future Research}

It should be apparent that our methods carry over without essential
modification for the analysis of more general
Landau-Ginzburg models, including
$(0,2)$ models.  The detailed analysis of the $\bar Q_{+,L}$ cohomology
can be more elaborate, but the principles are the same.  One
novelty (already known from the special case of Gepner models)
is that in general the number of massless $E_6$ singlets
at the Landau-Ginzburg ``point'' differs
from what it is in the field theory limit.

A number of interesting additional issues about these models
are worth pursuing.  In
particular, it should be possible to compute at least the
unnormalized Yukawa couplings; this would assist
in the investigation of real
phenomenology based on Landau-Ginzburg orbifolds.
It should also be straightforward to generalize our
approach to Landau-Ginzburg orbifolds with discrete torsion ~\Kenlg .

One of the most interesting prospects lies in the detailed
exploration of (0,2)
models.  Their rather complicated geometrical description makes them
hard to study by traditional techniques, but we have shown
that their Landau-Ginzburg description makes
them amenable to quite detailed
analysis.  One can write down (0,2) models with gauge groups
like $SO(10)$ or $SU(5)$, which are much less cumbersome
than $E_{6}$.  This makes (0,2) models perhaps the most
promising class of models for realistic phenomenology.
In addition, it is quite
plausible that a better understanding of (0,2) models
could lead to progress in the understanding of topology-changing
processes in string theory.

\bigskip
\centerline{\bf{Acknowledgements}}

We would like to thank J. Distler, N. Seiberg, and
A. Pasquinucci for helpful
discussions.

\listrefs
\end